\documentclass[aps,pra,twocolumn,showpacs,floatfix,superscriptaddress,nofootinbib]{revtex4}
\usepackage{graphicx}
\usepackage{hyperref}
\usepackage{amsmath}
\usepackage{amsfonts}

\def\>{\rangle}
\def\<{\langle}

\begin{document}

\title{Ontological Models for Quantum Mechanics and the Kochen-Specker theorem}

\author{Terry Rudolph}
\address{QOLS, Blackett  Laboratory, Imperial College London, Prince Consort Road, London SW7 2BW,
UK}
\address{Institute for Mathematical Sciences, Imperial College London, 53 Exhibition Road,
London SW7 2BW, UK}

\begin{abstract}

Certain concrete ``ontological models'' for quantum mechanics (models in which measurement outcomes are deterministic and quantum states are equivalent to classical probability distributions over some space of `hidden variables') are examined. The models are generalizations of Kochen and Specker's such model for a single 2-dimensional system - in particular a model for a three dimensional quantum system is considered in detail. Unfortunately, it appears the models  do not quite reproduce the quantum mechanical statistics. They do, however, come close to doing so, and in as much as they simply involve probability distributions over the complex projective space they do reproduce pretty much everything else in quantum mechanics.

The Kochen-Specker theorem is examined in the light of these models, and the rather mild nature of the manifested contextuality is discussed.
\end{abstract}

\maketitle

\section{Introduction}

If it is the case that orthodox quantum mechanics (QM) is an incomplete description
of nature, then several puzzling aspects of the theory become less so: The
role of conscious observers in the theory and collapse of the wavefunction
being the preeminent examples\footnote{For example, no one is paid to lie
awake at night puzzling about the role of conscious observers and `collapses' due to updating of information in classical
Liouville mechanics. Arguments in favor of this viewpoint have a long history, with early advocates such as Jaynes and Pierls. More recent arguments can be found in \cite{Spek1}. }. It is well appreciated that the Bell \cite{bell} and Kochen-Specker \cite{KS} theorems significantly constrain \emph{any} attempts at a more complete description. The KS theorem in particular is often taken as implying that any `complete' description of reality must be so strange that such attempts should automatically be abandoned. It seems to me, however, that the KS theorem tells us only that the outcomes of experiments must depend on the physical properties of both the measurement apparatus and the system under investigation - it provides evidence against the most dramatic form of reductionism. This raises the question: Without such reductionism at hand, must any complete description be so complicated that Occams razor necessitates its abandonment?  One purpose of this note is to show how a theory can be contextual without becoming particularly complicated, and without requiring a deep understanding of the dynamics of measurement processes in the theory.

I will be concerned here with the possible existence of what will briefly be termed \emph{ontological models} (OM's) - models which ideally reproduce the quantum predictions and have the following features:
Firstly, all the physical properties of an individual system are presumed to be
determined by (or determinable from) some mathematical object $\lambda$ (the ``ontic state'' or ``hidden variable''\footnote{I will avoid using the term `hidden variable' - in addition to being of bad repute in certain circles, some of the ontological models considered here have the same ``unhidden'' variables as regular quantum mechanics!} of the system)
which is an element of a space of such objects $\Lambda$. The quantum mechanical wavefunction $\left\vert
\psi \right\rangle,$ an element of a $d$ dimensional Hilbert space $\mathcal{H}^{(d)},$  is presumed an incomplete description of this underlying reality, and
thus corresponds in the OM\ to a probability density over $\Lambda:$
\begin{equation}\label{statecorrespondence}
\left\vert \psi \right\rangle \iff P_{\psi}\left(\lambda\right) d\lambda .
\end{equation}
Note that unlike the commonly utilized representations of quantum systems in terms of
quasi-probability distributions, which can become negative, this probability
density is presumed positive and normalized with respect to some
appropriate measure on $\Lambda$. Crucially, in OM's the probabilistic nature of quantum mechanical predictions arises \emph{only} from such classical uncertainty.

Secondly, measurement outcomes in OM's are deterministic. As such, if an $N$ outcome
measurement is performed which in QM is described by a (complete) set of orthogonal projection operators $\{\Pi_{i}\}_{i=1}^{N}$, then for a system in ontological state $\lambda $  there
is a unique one of these $N$ outcomes which will always be obtained. That is, if the value of $\lambda$ was known then the outcome which would be obtained is determinable. One can therefore divide
the space $\Lambda $ up into $N$ subspaces according to which of the
measurement outcomes each $\lambda\in\Lambda$ is associated with. Mathematically we define $N$
different \emph{characteristic functions} $\{\chi _{i}(\lambda )\}_{i=1}^{N}$
over $\Lambda ,$ which take the value 1 over the values of $\lambda $ which
yield the particular outcome, and which are 0 otherwise\footnote{For notational simplicity I am avoiding writing the characteristic function corresponding to $\Pi_i$ as $\chi_{i}(\lambda;\{\Pi_{j\neq i}\})$ which would more accurately reflect the possibility of contextuality, i.e. that the full set of measurement operators must be known in order to determine whether the system will give outcome $i$ or not.}. Thus the probability of
any particular measurement outcome is given by the correspondence:%
\begin{equation}\label{probcorrespondence}
\left\langle \psi \right\vert \Pi _{i}\left\vert \psi \right\rangle \iff
\int P_{\psi }(\lambda )\chi _{i}(\lambda )d\lambda .
\end{equation}

Finally, certain other assumptions necessarily lurk in the background of OM
descriptions of QM. The first is quite reasonable: What corresponds to the
Schr\"odinger evolution of the system is some sort of transformation of the
ontological state space $\Lambda $ into itself. The second is less simple:
Measurements must disturb the ontological state of the system. A non-disturbing measurement of some outcome $i$ would
result in the observer assigning the posterior distribution to the system:
\[
P_{i}(\lambda )d\lambda =\frac{\chi _{i}(\lambda )P_{\psi }(\lambda
)d\lambda }{\int P_{\psi }(\lambda )\chi _{i}(\lambda )d\lambda }.
\]%
However a sequence of such non-disturbing measurements would generally
allow the observer to narrow down their description of the system in such a
way that they obtain more predictability than we know to be the case for QM.
Thus it is presumed that upon measurement an imprecisely understood disturbance occurs which forces an observer to
``smear'' the post-measurement distribution they assign. Note that this smearing necessarily happens only
over states which lie in supp($\chi _{i}(\lambda )$), the support of the
characteristic function (i.e. the region over which the function is non-zero), because we know empirically that (in the absence of
intermediate evolution) a second measurement containing the same projector $\Pi _{i}$ must
yield this same outcome with certainty.


It is not known whether or not there exist OM's satisfying all of the above for an arbitrary dimensional quantum system. Kochen and Specker found such a model for a spin-1/2 (qubit) system - their model is reviewed in the next section. Subsequently I go on to examine similar models for a three dimensional (qutrit) system, models which unfortunately do \emph{not} exactly reproduce the quantum predictions. They do, however, come damn close. Since the KS theorem applies in three and higher dimensions it is interesting to see the manner in which these models manifest contextuality, as one suspects they must if they are to come as close to QM as they seem to do.

\section{Ontological model for a qubit}

\subsection{Kochen and Specker's ``Marble World''}

We begin with a simple and concrete model which
reproduces the operational predictions of the quantum mechanics of a qubit.  We then
formalize this model a little, which enables us to see how it can be
generalized to higher dimensional systems.

Our initial simple model, that we call ``marble world'', consists of a marble
constrained to move on the surface of a sphere. $\Lambda $ is then the set
of points of a unit sphere, and a generic element $\lambda $ can be parameterised
as a vector $[\sin \theta \cos \phi ,\sin \theta \sin \phi ,\cos \theta ]$
in the usual manner. Observers of marble world systems are somehow constrained -
they can never know the location of the marble exactly. In fact, the best
they can do is to assign a probability distribution $P_{\vec{n}}\left(
\lambda \right) $ over the space $\Lambda .$ Here $\vec{n}$ is a (fixed) unit vector which labels the distribution. Specifically,%
\begin{eqnarray*}
P_{\vec{n}}\left( \lambda \right)d\lambda  &=& (2(\lambda \cdot \vec{n})^2-1)d\lambda \mathrm{ \ \ \
if \ \ \ }\lambda \cdot \vec{n}\geq 0 \\
&=&0\mathrm{ \ \ \ \ \ \ \ \ \ \ \ \ otherwise}
\end{eqnarray*}%
Thus $P_{\vec{n}}\left( \lambda \right) $ is a cosine distribution peaked at
$\vec{n}$ and having support only on the hemisphere defined by $\vec{n}$ - it corresponds to the quantum mechanical pure state with Bloch vector $\vec{n}$. $d\lambda$ is the standard uniform measure over the sphere.

Measurements in marble world consist of the following: The sphere is observed from antipodal directions and the hemisphere in which the marble is actually located is noted. For concreteness imagine measurements are performed by means of different colored
lights (say red and green) being turned on above antipodal points $\vec{m},%
\vec{m}^{\perp }$ of the sphere. If the marble is in the hemisphere
illuminated by the green (resp. red) light it scatters a green (resp. red) photon which is
registered by the observer. It is not hard to verify that the probability of
this scattering is $\cos ^{2}\alpha $ $(resp. \sin^2\alpha)$ where $\alpha $ is the angle between $%
\vec{n}$ and $\vec{m}$ (resp. $\vec{m}^{\perp}$). Thus the quantum mechanical statistics (i.e. the Born rule) are easily reproduced. In the language of the introduction the characteristic function $\chi_{\vec{m}}(\lambda)$ is simply the function with value 1 over the hemisphere centered around $\vec{m}$ and 0 over the opposite hemisphere.

In order to make marble world correspond more closely to QM we
need to presume two other features: Firstly, we presume that a marble
scattering light is disturbed in such a way that after a measurement with outcome $\chi_{\vec{m}}$ the probability distribution an observer should assign the post-measurement
likelihood of the marble's location is simply $P_{\vec{m}}\left( \lambda \right)d\lambda.$
(E.g. the marble could be ``attracted'' to the light source it sees, and so the new
distribution becomes peaked at the point of greatest intensity, namely $\vec{m}$. Interestingly, in this particular example the intensity distribution of the light over the hemisphere will actually match the presumed cosine probability distribution!) Secondly we must presume the
dynamics in the theory takes the form of rotating the sphere ``underneath''
the marble so as to exactly mimic Schr\"odinger evolution. In this way marble world
exhibits all features of the physics of a single qubit.

\subsection{The SU(2) version of Kochen and Specker's model}

Because of the $SU(2)$ double covering of $SO(3)$ it is not hard to turn the
marble world description into a slightly more formal OM for a qubit, and it is this more formal model which is more easily generalized to higher dimensional systems.

The ontological state space $\Lambda $ consists of all rank 1 projectors
in $GL(2,\mathbb{C})$ (More precisely, it is the complex projective space upon which SU(2) acts transitively). The standard distance measure between $\lambda _{1},\lambda _{2}\in \Lambda $ (inducing the Frobenius norm) is
\[
d(\lambda _{1},\lambda _{2})=1-Tr\left( \lambda _{1}\lambda _{2}\right) .
\]%
Let $\lambda _{\psi }$ be the element of $\Lambda $ which is mathematically identical to
the matrix $\left\vert \psi \right\rangle\! \left\langle \psi \right\vert $
corresponding to quantum state $\left\vert \psi \right\rangle \in \mathcal{H}^{(2)}$
in some fixed (but otherwise arbitrary) basis. In the
ontological model the representation of a system in the quantum state $
\left\vert \psi \right\rangle$  is the probability density
\begin{eqnarray}\label{SU2PROB}
P_{\psi }(\lambda )d\lambda  &=&\mathcal{N}\left( Tr(\lambda\lambda _{\psi })-\frac{1}{2}\right) d\lambda \mathrm{ \ \ \ \ \ \ if \ \ }Tr(\lambda
\lambda _{\psi })\geq \frac{1}{2} \nonumber \\
&=&0\mathrm{ \ \ \ \ otherwise,}
\end{eqnarray}
where $\mathcal{N}$ is a normalization constant, and $d\lambda $ denotes a
suitable measure - in this case it is proportional to the standard $SU(2)$
invariant Haar measure. This distribution is  centered on (and peaked at) $\lambda _{\psi }$, and it has support over all elements of $\Lambda $
which are less than a distance 1/2 away from this point.

Since the model is deterministic, we need to specify for each $\lambda $ which outcome will be obtained when a measurement  is performed that is described in QM by von-Neumann projection operators $\{\Pi_{0}=|0\rangle\langle 0| ,\Pi_{1}=|1\rangle\langle 1|\}$.
If we denote by $\lambda _{0}$ and $\lambda _{1}$ the elements of $\Lambda $
which are identical to the quantum mechanical matrices $\Pi _{0},\Pi _{1}$
then we can define  characteristic functions over $\Lambda $:
\begin{eqnarray} \label{SU2CFN}
\chi _{0}(\lambda ) &=&\Theta \left( Tr(\lambda _{0}\lambda)-\frac{1}{2}\right) \\
\chi _{1}(\lambda ) &=&\Theta \left( Tr(\lambda _{1}\lambda )-\frac{1}{2}\right),
\end{eqnarray}
where $\Theta $ is the Heaviside step function. Heuristically then, a generic $
\lambda \in \Lambda $ gives the outcome to which it is closest. We refer to
the elements $\lambda _{0},\lambda _{1}$ as the ``central elements'' of the measurement outcomes, since they are at the centers of the characteristic functions, and moreover the ``decision'' a system makes about which outcome to give is determined with respect to these elements.

It is readily verified that with the above construction one obtains the exact correspondence
\[
\left\langle \psi \right\vert \Pi _{b}\left\vert \psi \right\rangle \iff
\int P_{\psi }(\lambda )\chi _{b}(\lambda )d\lambda ~~~~~~\mathrm{for \ }b=0,1.
\]
Again, in order to allow this ontological model to reproduce
the rest of familiar quantum mechanics for a qubit we must postulate that
any measurement disturbs the system in such a way that post-measurement an
observer's best posterior probability distribution is the one centred on $
\lambda _{b}$ (i.e. $P_{\Pi_b}(\lambda )$) according to the outcome $\chi _{b}
$ she obtained. We must also allow that coupling of the system to classical
fields causes an evolution in the ontic state space $\Lambda $ which amounts
to application of an $SU(2)$ rotation - as per Schr\"odinger evolution in QM.

Note that I have been very explicit about differentiating between the hidden variable space $\Lambda$ and the Hilbert space (or projective space) in which QM is formulated - despite the fact that mathematically for this particular OM they are identical spaces! This is partly for conceptual reasons, and partly because later I will look at describing a d-dimensional quantum system by an OM that is \emph{not} formulated over the d-dimensional Hilbert space, but rather is formulated in a higher dimensional Hilbert space.

It has been known since the seminal work of Kochen and Specker that
ontological models for systems which are described in QM by Hilbert spaces of
dimension greater than two must be \emph{measurement contextual.} The KS model\ for a qubit is actually trivially non-contextual. This can be seen from the expressions for the characteristic functions (\ref{SU2CFN}): a system in an ontic state $\lambda$ can tell whether or not it is going to ``give'' the outcome $\Pi_0$ (say) without having to know what other measurement outcomes (of course there is only one!) are being measured simultaneously with $\Pi_0$.

I will now describe a very similar model for a qutrit, and then in the next section
examine how measurement contextuality is manifested in the model.

\section{Qutrit model I}

For this qutrit model the ontological state space $\Lambda $ consists of all rank 1 projectors in $GL(3,\mathbb{C}).$ As above, the distance measure between $\lambda _{1},\lambda _{2}\in
\Lambda $  is
\[
d(\lambda _{1},\lambda _{2})=1-Tr\left( \lambda _{1}\lambda _{2}\right) .
\]
and the representation of $\left\vert \psi \right\rangle $ (now an element
of $\mathcal{H}^{(3)}$) in the ontological model is the probability density
\begin{eqnarray}\label{qutritprob}
P_{\psi }(\lambda )d\lambda  &=&\mathcal{N}\left( Tr(\lambda\lambda_{\psi })-\Delta\right) d\lambda \mathrm{ \ \ \ \ \ \ if \ \ }Tr(\lambda
\lambda _{\psi })\geq \Delta \nonumber\\
&=&0\mathrm{ \ \ \ \ otherwise,}
\end{eqnarray}
where $\mathcal{N}$ is a (different) normalization constant and $d\lambda $
is now proportional to the $SU(3)$ invariant Haar measure. Note that a new parameter $\Delta$ has been introduced, which essentially determines how large the support of the probability distribution corresponding to the quantum state is.

When a measurement that is quantum mechanically described by von-Neumann
projection operators $\{\Pi _{0}=|0\>\<0|,\Pi _{1}=|1\>\<1|,\Pi _{2}=|2\>\<2|\}$ is
performed then a system in a generic state $\lambda $ gives
the outcome according to which of the corresponding central elements $\lambda
_{0},\lambda _{1},$ or $\lambda _{2}$ it is closest to. More formally, such
a measurement is described by the characterisitic functions:
\begin{eqnarray*}
\chi _{0}(\lambda ) &=&\Theta\left( Tr(\lambda\lambda_0)-Tr(\lambda\lambda_1) \right)
\Theta\left(Tr(\lambda\lambda_0)-Tr(\lambda\lambda_2)\right) \\
\chi _{1}(\lambda ) &=&\Theta\left( Tr(\lambda\lambda_1)-Tr(\lambda\lambda_0) \right)
\Theta\left(Tr(\lambda\lambda_1)-Tr(\lambda\lambda_2)\right) \\
\chi _{2}(\lambda ) &=&\Theta\left( Tr(\lambda\lambda_2)-Tr(\lambda\lambda_0) \right)
\Theta\left(Tr(\lambda\lambda_2)-Tr(\lambda\lambda_1)\right).
\end{eqnarray*}

\begin{figure}
{\includegraphics[width=8cm]{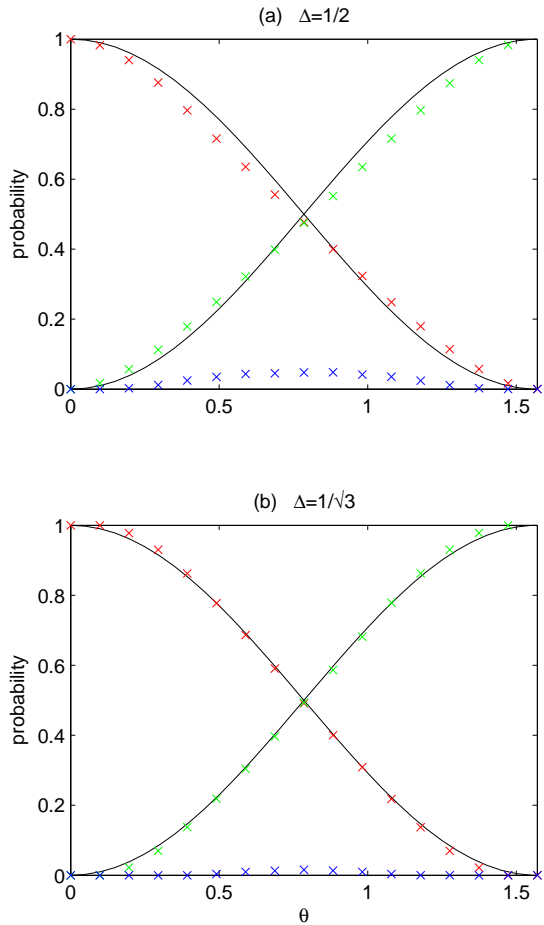}}
\caption{ \label{SU3cosqd} Comparison between OM and QM predictions for the first qutrit model; the two figures are for different $\Delta$. The quantum state in the measurement basis is $|\psi\>=[\cos\theta,\sin\theta,0]$, thus the QM predctions for outcomes 0 and 1 are $\cos^2\theta,\sin^2\theta$. The blue crosses correspond to the probability of the $\Pi_2$ outcome which in QM is exactly 0.}
\end{figure}

It is not simple to analytically examine the extent to which the desired correspondence of Eq.~(\ref{probcorrespondence}) is satisfied, because of the complexity of parameterizing the integration regions.\footnote{I should point out that the paper \cite{tilma} has been
particularly helpful for obtaining explicit parameterizations
of the group manifold for the small amount of analytic work I have done.} As such I simply do the integrations numerically by a monte-carlo method.
It turns out that this OM does \emph{not} reproduce the quantum statistics perfectly - i.e. the correspondence of Eq.~(\ref{probcorrespondence}) is not exactly satisfied. Empirically I find that the maximum deviation from the quantum prediction occurs when the quantum state lies in a 2-dimensional subspace spanned by two of the three considered measurement outcomes. That is, taking $|\psi\>=\cos\theta|0\>+\sin\theta|1\>$ and then examining the probabilities for obtaining outcomes $\{\Pi _{0}=|0\>\<0|,\Pi _{1}=|1\>\<1|,\Pi _{2}=|2\>\<2|\}$ gives by far the largest deviations from the quantum predictions. These deviations can be succinctly plotted as in  Fig.~\ref{SU3cosqd}. In the figure the black lines correspond to the standard quantum mechanical $\cos^2\theta$ and $\sin^2\theta$ predictions for outcomes $\Pi_0$ and $\Pi_1$. The red and green crosses are the corresponding predictions of the OM. Note that while the probability of the $\Pi_2$ outcome is strictly 0 in QM, in the OM it is finite (the blue crosses). The figure shows the curves for two different values of $\Delta$. There is nothing particularly special about $1/\sqrt{3}$ - I have not had the patience (the monte-carlo integrations take quite a while) to find the actual optimal choice, but trying a few values and comparing ``by eye'' it seems close to the optimal.

Note that the reason the probability of obtaining the $\chi_2$ outcome is non-zero in the OM is that whenever $\Delta<2/3$ there exist $\lambda$ which are closer to $\lambda_2$ than $\lambda_0,\lambda_1$ but which are still close enough to $\lambda_\psi$ (for $\theta\approx\pi/4$) to lie within the support of $P_\psi(\lambda)$. Why then not choose $\Delta=2/3$? Doing so seems to cause the other curves to deviate more from the quantum predictions. There are a large variety of ``tweaks'' of this model that one can imagine, and I do not have the computing power (or patience) to find optimal parameter choices for each such tweak.

Before I go on to describe a different OM that I personally find more appealing than this first qutrit version, I am going to digress a little and talk about the Kochen-Specker theorem and this particular OM. This is because although this model does not exactly reproduce QM, I believe the sort of manner in which it manifests contextuality is quite natural and much more benign than the dramatic statements one often sees about the implications of the Kochen-Specker theorem for realist interpretations of quantum mechanics might lead one believe.

\section{The Kochen-Specker Theorem}

One of the most profound restrictions on interpretations of QM is
provided by the Kochen-Specker theorem, which holds for three or more
dimensional quantum systems. The OM of the previous section exhibits the
contextuality required by the KS theorem in the following way. Consider two
different complete (projection-valued) measurements in QM: $\{\Pi _{0},\Pi _{1},\Pi
_{2}\}$ and  $\{\Pi _{0},\Pi _{1}^{\prime },\Pi _{2}^{\prime }\}.$ Recall
that our rule assignment to an arbitrary element $\lambda $ is that a system
in state $\lambda $ gives the outcome corresponding to the central element
to which it is closest. Now, there exist some ontological states $\lambda $ that are closer to $\lambda _{0}$ than to $\lambda _{1},\lambda _{2},$ but which are closer to one
of $\lambda _{1}^{\prime },\lambda _{2}^{\prime }$ than to $\lambda _{0}.$

This can be seen simply by means of the 3 dimensional real space analogue in Figure 1, which loosely corresponds to the $\Delta=1/2$ version of the qutrit OM described above. In the figure the
three orthogonal unit vectors represent  $\lambda _{0}$, $\lambda _{1}$ and $\lambda
_{2}$, and the quadrant of the sphere depicted is part of the space $\Lambda$. The red lines depict the boundaries of the supports of the
characteristic functions  $\chi _{b}(\lambda )$ - that is, elements $\lambda$ within the red boundary are closest to the appropriate vector $\lambda_b$.  Clearly if the space is
rotated around $\lambda _{0}$ to obtain a new set of vectors $\lambda _{0},\lambda
_{1}^{\prime },\lambda _{2}^{\prime }$ then those points $\lambda \in \Lambda $ which
lie in the region marked ``unfaithful'' will change the outcome they are
associated with from $\lambda _{0}$ to either $\lambda_{1}^{\prime}$ or $\lambda_{2}^{\prime}$.  We
will refer to elements $\lambda $ which always give outcome $\lambda _{0}$
as ``faithful'' to $\lambda _{0}.$

Thus the measurement outcomes in the OM are contextual: sometimes (interestingly, however, not all the time) the system is in an ontic state $\lambda$ for which knowledge of all three measurement outcomes being measured simultaneously is required in order to determine which outcome will be obtained. That is, the probability of a specific outcome depends on its context.

\begin{figure}
{\includegraphics[width=6cm]{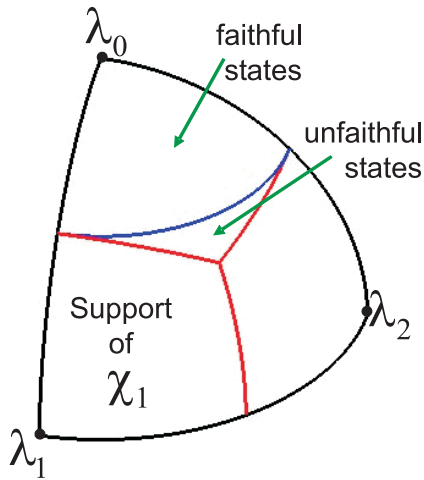}}
\caption{ \label{OMfigure} A real space analogue depicting a quadrant of the space $\Lambda$ and three orthogonal ``central elements'' $\lambda_0,\lambda_1, $ and $\lambda_2$. The regions of ontic states closest to each such element are bounded by the red curves. The ontic states which always give the $\Pi_0$ outcome regardless of their context are indicated as ``faithful''. The set of faithful states also forms $supp(P_0(\lambda))$ for the particular choice of $\Delta=1/2$.}
\end{figure}

I now proffer the opinion that the contextuality of this OM does not, however, encumber us with the conceptual problems which are often associated with understanding the implications of the Kochen-Specker theorem. Roughly speaking, these puzzling implications are usually taken to be that the theorem presents an obstruction to our making sense of the system ``having values'' of certain physical quantities which are subsequently revealed by measurements.  The OM sidesteps these conceptual issues because it is, in some sense, \emph{relational}. More precisely, we should expect that a more complete description of measurements would involve consideration of the ontic states associated with the measurement apparatus, and it is not unreasonable that the response of the system to a measurement depends on the \emph{relationship} between the ontic state of the apparatus and the ontic state of the system. Because the measurements  $\{\Pi _{0},\Pi _{1},\Pi_{2}\}$ and  $\{\Pi _{0},\Pi _{1}^{\prime },\Pi _{2}^{\prime }\}$ involve different physical arrangements, we should expect that the corresponding ontic states of the measurement apparatus itself are different. Without the more complete description we do not (yet) have a proper description of the apparatus' ontic states. However, since in our OM the measurement outcome is determined solely by the relationship between the \emph{central elements} and the actual state $\lambda$ of the system, it is quite uncontroversial to view the central elements $\{\lambda_0,\lambda_1,\lambda_2\}$ or $\{\lambda_0,\lambda'_1,\lambda'_2\}$ as in a natural correspondence with these (as yet undetermined) apparatus ontic states.

Now, physics would be an extremely difficult endeavor if we needed to know everything about the measurement apparatuses \emph{and} the systems under investigation to form a coherent understanding of the world - in effect we need the simplifications afforded by the implicit reductionism which allows us to talk about systems at all. The KS theorem denies us the most  extreme form of reductionism. However, in the sort of OM discussed here we obtain the next best thing: we do not need to know everything about the highly complex physical reality of the macroscopic measurement devices - we need only know that the important features of the apparatus for the purposes of understanding its role as a measuring device can be encapsulated in the simple mathematical objects $\{\lambda_i\}$, which form the central elements of the measurement.

One upshot of this is that the quantum projector $\Pi _{0}$ is always associated with a \emph{unique} central element $\lambda_0$. In \cite{Spek2} Spekkens showed how contextuality could be defined operationally (and thus for any physical theory, not just quantum mechanics) in terms of the redundancies in the mathematical description of one theory by another. Such mathematical redundancy is certainly present at the level of characteristic functions in this OM: there are a continuous infinity of characteristic functions $\chi_0(\lambda)$ corresponding to $\Pi_0$, one for each of the continuous infinity of possibly co-measured projectors (corresponding to the different rotations about the $\lambda_0$ axis in Fig.~\ref{OMfigure}). If we quantify contextuality by simply counting this redundancy, then this OM is highly contextual. In my opinion, however, the contextuality of the model is ameliorated by the fact that there is no need to have a large mathematical redundancy in the ontic states $\{\lambda_0,\lambda_1,\lambda_2\}$, the relations to which represent ``actual physical quantities''. Looking at this another way, there is a simplicity in the \emph{mathematical rules} from which the contextual measurement outcomes are determinable. It is not as if we have to compile a huge table of different outcomes for each of the different possible measurements and all possible input ontic states. It seems likely to me that at the level of just counting `representation redundancies' all OM's will need to be infinitely (preparation and measurement) contextual\footnote{Though this is not easy to prove - just ask Nicholas Harrigan!}. Some models will manifest that contextuality in a more simple and natural way than others.

A final word about relationalism. There is a large body of work advocating identification of relational ``elements of reality'' in classical physical theories - General Relativity in particular. Even in terms of non-relativistic QM, however, it should be realized that the quantum mechanical wavefunction is itself devoid of operational meaning unless the experimentalist understands which external devices the relative phases in a particular superposition are defined with respect to. It is perhaps not too much of a stretch to see contextuality in general as simply a reflection of this essential feature of physics. The puzzle then becomes why it is that classical physics is non-contextual.

%
%
%
%
%

%

\section{Ontological models with uniform probability densities}

%

Upon first exposure to the Kochen-Specker qubit model of section II my initial reaction was to feel slightly cheated with regard to the manner in which the Born rule is obtained. That is, the specific choice of a cosine distribution peaked at the corresponding quantum mechanical Bloch vector is clearly what leads directly to the quantum mechanical ``$\cos^2$'' type of probability distribution.

If some sort of OM really does underpin QM, then we should expect that the probability density would take some ``natural'' form. In fact there are very few probability densities which arise naturally in physics - and the ones which do (such as the Gibbs distribution) tend to be understandable in terms of an application of a maximum entropy principle. In looking for OM's there is a further intuition at play - namely that many of the strange features of quantum mechanics are derivable from us as observers being subject to an encompassing \emph{epistemic restriction} (See e.g. \cite{Spek2} for much more on this sort of thing). It seems intuitive that such restrictions would generally lead to assignment of \emph{uniform} probability distributions over the `hidden' variables. (Again, see \cite{Spek2} for this type of thing at play).

There is another point in favor of looking for uniform distributions: As has been mentioned several times, `collapse' - viewed  as an updating of knowledge in an OM - is still necessarily accompanied by physical disturbance. Collapses which induce a ``disturbing dynamics'' that then leads to a new assignment of some strange probability density function would presumably have to be quite specialized. Given the variety of systems to which QM applies this seems somewhat unlikely. Thus I have spent some time looking for OM's which, like the above models, preserve much of the Hilbert space structure of QM, but which don't require assigning some sort of contrived probability density in order to conform closely to QM.\footnote{The preservation of as much of the Hilbert space structure is important: for example, it means that I do not have to look for new Hamiltonians (or some new equivalent thereof) - unitary evolution can still be the basic form of evolution in the OM, and the spectrum of hydrogen can go unchanged...}

Here is one example of an OM which only involves uniform distributions. We let the quantum system which has dimension $d$ be described in an ontic space $\Lambda$ consisting of
all rank 1 projectors in $GL(d_\Lambda,\mathbb{C}),$ where $d_\Lambda>d$. (In fact for the examples presented here I'll take $d_\lambda=d+1$.) Corresponding to the quantum state $|\psi\>\<\psi|$ there is an element $\lambda_\psi\in \Lambda$ defined via:
\[
\lambda_\psi=\left[\begin{array}{cccc}|\psi\>\<\psi| & 0\\ 0&0\end{array}\right]
\]
(i.e.  the bottom right block is $(d_\Lambda-d) \times (d_\Lambda-d)$ dimensional). This is simply choosing $\lambda_\psi$ exactly equal to $|\psi\>\<\psi|$ on a fixed d-dimensional subspace of $\Lambda$. Elements $\lambda_i$ corresponding to the QM measurement projectors $\Pi_i$ can similarly be defined.

\begin{figure}
{\includegraphics[width=8cm]{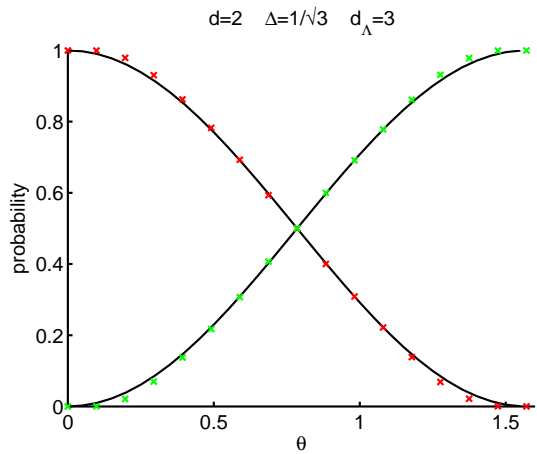}}
\caption{ \label{SU2_in_d3_uniform} The probabilities of obtaining $\Pi_0=|0\>\<0|,\Pi_1=|1\>\<1|$ for a qubit in state $|\psi\>=\cos\theta|0\>+\sin\theta|1\>$. The black curves are the QM predictions, the crosses are the probabilities for an OM based on uniform probability distributions in a 3-dimensional ontic state space. }
\end{figure}

\begin{figure}
{\includegraphics[width=8cm]{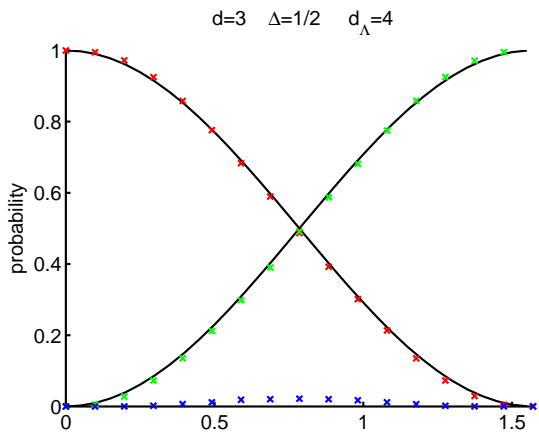}}
\caption{ \label{SU3_in_d4_uniform} An equivalent plot to Fig.~\ref{OMfigure}, but now for  a qutrit OM based on uniform probability densities in a 4-dimensional ontic state space.}
\end{figure}

The probability density corresponding to quantum state $|\psi\>$ is then
\begin{eqnarray}\label{qutritprob}
P_{\psi }(\lambda )d\lambda  &=&\mathcal{N} d\lambda \mathrm{ \ \ \ \ \ \ if \ \ }Tr(\lambda
\lambda_{\psi })\ge \Delta \nonumber\\
&=&0\mathrm{ \ \ \ \ otherwise,}
\end{eqnarray}
Note that I am leaving in the parameter $\Delta$ which determines the size of the support of $P_{\psi }(\lambda )d\lambda$.

The rule for determining measurement outcomes (and therefore the characteristic functions) are unchanged from that of the qutrit model discussed previously.

In Fig.~(\ref{SU2_in_d3_uniform}) the the quantum probabilities and those obtained for the $d=2$ version of this OM are plotted. Unlike the Kochen-Specker OM for a qubit this one does not reproduce the quantum probabilities exactly. Fig.~(\ref{SU3_in_d4_uniform}) is a plot for the qutrit version of this OM - it essentially does no worse than the original qutrit OM despite it utilizing only uniform distributions. The interesting conclusion is that something close to the desired sin/cos distributions can be ``inherited'' in what I like to think of as a kind of `concentration of measure' effect (though I'm not sure this is exactly what is going on, it is simply where the intuition came from).

\section{What do we know about general ontological models?}

The Kochen-Specker theorem is essentially the only nontrivial theorem known which affects any attempt at constructing an ontological model (taking the viewpoint that locality is simply a way of enforcing non-contextuality in a natural manner, and Bell's theorem demonstrates the need for a violation of this).

For each quantum mechanical projector $\Pi _{0}=|0\>\<0|,$ a system prepared in $\left\vert
0\right\rangle $ gives this outcome with probability 1, and in the OM's presented here this
is reflected in the fact that $\int P_{0}(\lambda )\chi _{0}(\lambda
)d\lambda =1.$ Clearly, however, $P_{0}(\lambda )$ must have support only
over states faithful to $\lambda _{0}.$ If not, then a system  prepared
in $P_{0}(\lambda )$ would sometimes give an outcome different to $\chi _{0}
$, were an observer to vary over the multiple sets of measurements containing this projector/central element. We see therefore that in the OM's considered here the strict inclusion holds:
\begin{equation*}
\mathrm{supp}\left( P_{0}(\lambda )\right) \subset \mathrm{supp}\left( \chi
_{0}(\lambda )\right) .
\end{equation*}
(In the analogy of Figure 1 the boundary of the support of $P_{0}(\lambda )$ for $\Delta=1/2$ is denoted by the blue line of latitude - i.e the probability distributions have support over the complete set of states faithful to their associated measurement.)

One may well ask whether this strict inclusion is a feature specific to these ontological models. In some related work \cite{harr2} we show that in fact
\emph{every} reasonable ontological model of three or more dimensional quantum systems \emph{must} have this feature for an infinite number (though perhaps not all) of the probability distributions corresponding to quantum states.  Interestingly, this is proven by applying a Kochen-Specker style
construction to quantum \emph{preparation procedures} instead of measurements.

\section{Conclusions}

It would be nice to find an OM which is so close to the quantum predictions that current experiments cannot rule it out. (I am certainly interested in hearing about present experimental bounds on how precisely the Born rule is known to be satisfied in higher dimensional systems - this can then give me a target to `shoot for'.) On prime numbered days I am not even sure current experiments rule out the OM's discussed above. I have played around with many variations on the sorts of OM I have discussed here that use different spaces $\Lambda$, but have not yet found an OM reproducing the quantum statistics exactly for any dimension $d$. I know of no proof, however, that such an OM cannot be lurking out there. I certainly hope that one is -- at present it seems the only way QM can really make sense to me.

\begin{acknowledgements}
This paper would not have arisen without my many stimulating and conceptually important discussions with Nicholas Harrigan, who declined an offer to share authorship of it. Over the years my thoughts about an ontology for quantum mechanics have benefited greatly through discussions with Rob Spekkens.
\end{acknowledgements}


\section*{References}
\begin{thebibliography}{99}
\bibitem{bell} J. S. Bell, Rev. Mod. Phys. \textbf{38}, 447 (1966).
\bibitem{KS} S. Kochen and E. P. Specker, J. Math. Mech. \textbf{17}, 59 (1967).
\bibitem{Spek1} R.W. Spekkens, eprint:quant-ph/0401052.
\bibitem{fuchs} C.A. Fuchs, quant-ph/0105039.
\bibitem{tilma} T. Tilma and E.C.G. Sudarshan, J. Phys. A: Math. Gen. \textbf{35} 10467, (2002). Also at eprint:math-ph/0205016v5.
\bibitem{Spek2} R.W. Spekkens, Phys. Rev. A \textbf{71}, 052108 (2005). Also at eprint:quant-ph/0406166.
\bibitem{harr2} N. Harrigan, T. Rudolph and R.W. Spekkens (in preparation).
\end{thebibliography}
\end{document}